\documentclass[showpacs,twocolumn,aps,preprintnumbers,letterpaper,longbibliography]{revtex4-1}
\usepackage{amsmath,amssymb}
\usepackage{epsfig}
\usepackage{graphicx}
\usepackage{bm}
\usepackage{amsmath}
\usepackage{slashed}
\usepackage{amsfonts}
\usepackage{epstopdf}
\usepackage[colorlinks,linkcolor=blue,citecolor=blue]{hyperref}
\newcommand{\be}{\begin{equation}}
\newcommand{\ee}{\end{equation}}
\newcommand{\bear}{\begin{eqnarray}}
\newcommand{\eear}{\end{eqnarray}}
\newcommand{\ba}{\begin{array}}
\newcommand{\ea}{\end{array}}

\def\be{\begin{eqnarray}}
\def\ee{\end{eqnarray}}
\def\bea{\be}
\def\eea{\ee}

\def\roughly#1{\mathrel{\raise.3ex\hbox{$#1$\kern-.75em%
\lower1ex\hbox{$\sim$}}}}

\begin{document}

\title{Mass structure of hadrons and light-front  sum rules in $^\prime$t Hooft model}

%\author{Xiangdong Ji$^{1}$, Yizhuang Liu$^{2}$  and Ismail Zahed$^3$}
%\email{xji@physics.umd.edu}
%\email{yizhuang.liu@sjtu.edu.cn}
%\email{ismail.zahed@stonybrook.edu}
%\affiliation{$^1$Department of Physics, University of Maryland, College Park, Maryland 20742, USA\\
%$^2$ Tsung-Dao Lee Institute, Shanghai Jiao Tong University, Shanghai, 200240, China\\
%$^3$ Department of Physics and Astronomy, Stony Brook University, Stony Brook, New York 11794-3800,USA}

\author{Xiangdong Ji}
\email{xji@physics.umd.edu}
\affiliation{Department of Physics, University of Maryland, College Park, Maryland 20742, USA}

\author{Yizhuang Liu}
\email{yizhuang.liu@sjtu.edu.cn}
\affiliation{Tsung-Dao Lee Institute, Shanghai Jiao Tong University, Shanghai, 200240, China\\
 Institut  fur Theoretische Physik, Universitat Regensburg, D-93040 Regensburg, Germany}

\author{Ismail Zahed}
\email{ismail.zahed@stonybrook.edu}
\affiliation{Center for Nuclear Theory, Department of Physics and Astronomy, Stony Brook University, Stony Brook, New York 11794--3800, USA}

%%%%%%%%%

\date{\today}
\begin{abstract}
%We analyze  the role of the vacuum condensates in the light cone distributions of mesons using two-dimensional spinor QCD.
We study the mass/energy structure of the bound state of hadrons in two-dimensional quantum chromodynamics
 in the large number of color limit  ($^\prime$t Hooft model). We analyze  separately the contributions from the
traceless and trace part of the energy-momentum tensor, and show that the masses are related to the matrix elements
of the scalar charge and Coulomb energy. We derive the light-front sum rules for the scalar charge
and Coulomb energy, expressed in terms of the light-front wave functions, and find that they are
regular at $x=0$ without the delta function contribution. We also consider the result
for the massless Goldstone boson, as well as the structure of the gravitational form factors of the
bound meson states.
\end{abstract}
%\pacs{11.25.Tq, 11.15.Kc, 12.38.Lg}
%\pacs{11.25.Tq, 13.60.Hb,13.85.Lg}
%11.25.Tq 	Gauge/string duality  13.60.Hb for deep-inelastic structure functions; 13.85.Lg 	Total cross sections

%11.15.Kc	Classical and semiclassical techniques
%11.30.Rd	Chiral symmetries
%12.38.Lg	Other nonperturbative calculations

\maketitle

\setcounter{footnote}{0}

%\baselineskip 18pt \pagebreak
%\renewcommand{\thepage}{\arabic{page}}
%\tableofcontents
%\pagebreak

\section{Introduction}

One of the chief driving force for the future of the Electron-Ion collider (EIC) is the
understanding of the origin of the nucleon mass~\cite{Report}.
There has been a lot of interest in understanding the mass structure following the original
approach in Refs. \cite{Ji:1994av,Ji:1995sv} by studying the matrix elements of the energy-momentum tensor
of quantum chromodynamics (QCD). A recent review can be found in \cite{Metz:2020vxd}, see also
\cite{Gryniuk:2020mlh,Hatta:2020iin,Borsanyi:2020bpd,Zeng:2020coc,Chen:2020gml} for new
developments.
It is widely accepted, that its origin
stems from the anomalous breaking of scale-invariance or the trace anomaly in quantum chromodynamics
(QCD) which generates  a non-perturbative mass scale, that is likely tied to the confinement of color
and the spontaneous breaking of chiral symmetry in the QCD dynamics. A measure of the trace anomaly
is captured by the matrix element of the gluon operator
$F^2$, which is related to the twist-four gluon distribution function on the light cone~\cite{Ji:2020baz}.
Its first moment or the light-cone sum rule is related to the gluon matrix element in a nucleon state.

%The partonic distributions are inherently non-perturbative. They are currently accesses through high-energy
%experiments, lattice simulations or models.
%Recently one of us~\cite{JI1} has suggested that the light cone hadronic wavefunctions
%can be recovered from Euclidean correlators in hadronic states using instead
%quasi-parton distribution functions  through pertinent renormalization in the infinite momentum limit.
%Preliminary lattice simulations have proven very promising~\cite{JI2,ALEX}. For a recent review of this
%approach and results we refer to~\cite{Ji:2020ect} (and references therein).

The purpose of this letter is to shed some light on the origin of mass in two-dimensional spinor QCD, where fermionic
and gluonic condensates develop in the vacuum state in the weak coupling regime and large $N_c$~\cite{tHooft:1974pnl},
following on earlier investigations of the quasi-distributions~\cite{Jia:2018qee,Ji:2018waw} and light-front
mass sum rule~\cite{Ji:2020baz}.
In passing, we will briefly quote the result for two-dimensional QED.
We will construct the corresponding chiral-odd quark and twist-four gluon distribution functions,
and discuss their moments in relationship to the contribution to the mass/energy
of the hadrons.

The organization of the paper is as follows: in section~\ref{INTRO} we provide an introduction to the model and its quantization in large $N_c$ limit. Both the equal-time quantization formalism and the light-front quantization formalism are introduced. In section~\ref{MASSS}, we introduce the symmetric energy-momentum tensor for
two-dimensional QCD and discuss the mass sum rule in the rest frame.
In section~\ref{DISTRIBUTION} we discuss the mass sum rule in the infinite momentum frame (IMF), and relates the quark and gluon contributions to light-front distribution functions. More specifically, in~\ref{QUARK} we derive the chiral odd quark parton distribution function, and show that its zeroth moment is the scalar quark condensate or sigma term in a bound meson without  the presence of a delta-function.
We also show that the vacuum chiral condensate is directly expressible in terms of the would-be Goldstone parton distribution function.
In~\ref{GLUON} we derive the twist-four gluon parton distribution function in a bound meson state, and analyze its zeroth moment.
In section~\ref{GRAVITON} we show that the energy momentum tensor in the bound mesons is characterized
by two-invariant form factors in relation to the graviton and dilaton form factors, suggesting the possibility of a direct measurement of  the trace part of the energy-momentum tensor entering the mass sum rule in diffractive processes.  In section~\ref{ANOMALYX} we derive a dual mass sum rule  with the help of
the colored axial-vector anomaly, where the mass budget is solely fermionic in any frame.
%we show how the colored axial-vector  anomaly yields a subdominant quantum  screening of the electric field, and how this screening allows a recast of the mass sum rule in terms of dominant local 4-Fermi interactions in the large $N_c$ limit.
Our conclusions are in section~\ref{CONCLUSIONS}. 
We briefly quote the results for  massive two dimensional QED in Appendix~\ref{QED}. Some additional aspects of the light cone wavefunctions in two dimensions are detailed in Appendix~\ref{LIGHT},
and a general Virial theorem is derived in Appendix~\ref{virial}.

%Two-dimensional scalar QCD has a smooth large $N_c$ limit with a confining spectrum~\cite{TSAO,TAM,GRIN}. In this model the current correlators exhibits many features of four-dimensional QCD in contrast to two-dimensional spinor QCD~\cite{HOOFT}. In the deep inelastic  regime  the results exhibit expected scaling laws, and are overall in support of the Feynman partonic picture and the light cone expansion. In this paper, these two models will be used interchangeably to test the concept of the quasi-distributions in a non-perturbative context, as they differ by a minor change in the algebra of the pertinent bosonic operators. Specifically, we  construct the quasi-parton distributions for both scalar and spinor QCD in leading and subleading order in $1/N_c$ and show that they merge with the expected light cone distributions in the infinite momentum limit without additional renormalization. Our leading conclusion for two-dimensional spinor QCD is in agreement with a recent study~\cite{RECENT}.

\section{Introduction to the Model}\label{INTRO}
In this section we introduce the model, its quantization  and its solution in the large $N_c$ limit. The Lagrangian of the theory is

\begin{align}
{\cal L}=-\frac{1}{4}F^{a\mu\nu}F^a_{\mu\nu}+\bar \psi(i\gamma\cdot D-m)\psi \ .
\end{align}
Our space-time coordinates reads $(t,z)$, the metric is $(1,-1)$. We choose the Weyl-like  basis
$\gamma^\mu=(\sigma^t,-i\sigma^z)$ with $\gamma^5=\gamma^t\gamma^z=\sigma_3$ where $\sigma_i$ are traditional Pauli matrices. The covariant-derivative reads $D^\mu=\partial^\mu-ig_0A^{a\mu}T^a$, modulo regularization. The color matrices are normalized to ${\rm Tr}(T^aT^b)=\delta^{ab}/2$ with $a=1, .., N_c^2-1$ and $\psi$ is in the fundamental representation. In two-dimensions, the gluonic contribution can be simplified using the observation that the anti-symmetric field strength relates to a gauge-covariant and colored pseudo-scalar potential $E^a$

\be
\label{2}
F^{a\mu\nu}=\epsilon^{\mu\nu}E^a \ .
\ee

In two dimensional QCD, unlike the four dimensional  version, the gauge-coupling $g_0$ has  mass dimension equal to $1$.  The natural mass-scale in the gauge sector is setup by $m_0^2=\frac{g^2N_c}{2\pi}\equiv \frac{\lambda}{2\pi}$. For pure Yang-Mills  in two dimensions, no dynamical degree of freedom such as the glueballs can be generated. The only intensive quantity with natural physical meaning is the string tension $\sigma=\pi m_0^2/2$.  When  fermions are  included, there is one more mass scale, the bare fermion mass $m$. The confining potential between color charges allows the formation of mesonic bound state, which can be exactly solved in the large $N_c$ limit.  The bound-state equation for the mesons can either be derived in the $A^z=0$ axial gauge using equal-time quantization, or in the light-cone gauge using light-front quantization. It can be shown that by imposing the large meson momentum $P^z\rightarrow \infty$ limit on the equal-time wave function, one recovers the light-front wave functions~\cite{Jia:2018qee,Ji:2018waw}.

\subsection{Equal-Time Quantization in $A^z=0$ Gauge.}
We first introduce the equal-time quantization in the axial gauge, which was  investigated first in Ref.~\cite{Bars:1977ud}. See also Ref.~\cite{Jia:2018qee} for a nice introduction. In this gauge, the equation of motion for the temporal component of the gauge field is non-dynamical and can be expressed in terms of quark fields

\begin{align}
A^{0a}=\frac{1}{\partial_z^2}g_0\bar \psi \gamma^0T^a\psi \ .
\end{align}
Thus, the pseudo-scalar $E^a$ can be written as

\begin{align}
E^a=-\frac{1}{\partial_z}g_0\bar \psi \gamma^0T^a\psi  \ .
\end{align}
The Hamiltonian reads in terms of the quark field

\begin{align}\label{eq:H}
H= &\int dx \bar \psi\left(-\gamma^z\partial_z+m \right)\psi +\nonumber \\
&\frac{g_0^2}{2}\int dzdz' (\bar \psi \gamma^0T^a\psi)(z)\frac{1}{-\partial_z^2}(z-z')(\bar \psi \gamma^0T^a\psi)(z') \ .
\end{align}
To diagonalize the Hamiltonian, let us introduce the dressed quark field

\begin{align}
\psi_A(z)=\int\frac{dp}{2\pi}\frac{1}{\sqrt{2E(p)}}e^{ipz}\bigg(u(p)a_A(p)+v(-p)d_A^{\dagger}(-p)\bigg)
\end{align}
with $A=1,..,N_c$. The anti-commuting relations for the creation-annihilation operators are

\be
\label{COM}
\left[a_A(p),a_B^\dagger(p')\right]_+&=\left[b_A(p),b_B^\dagger(p')\right]_+\nonumber \\
&=\delta_{AB}(2\pi)\delta(p-p') \ .
\ee
By writing the spinors as

\begin{align}
u(p)&=\sqrt{E(p)}e^{-\frac{\gamma^z}{2}\theta(p)}(1,1)^T \ ,\\
v(-p)&=\sqrt{E(p)}e^{-\frac{\gamma^z}{2}\theta(p)}(1,-1)^T \ ,
\end{align}
and plugging into the Hamiltonian, one finds that there is  no linear term in the creation-annihilation operators $a_A$, $b_A$ if $E(p)$ and $\theta(p)$ satisfy the gap equations

\begin{eqnarray}
&&E(p)=m\cos \theta(p)+p\sin \theta(p)+\frac{m_0^2}{4}\int_{\rm PV} \frac{\cos[\theta(p)-\theta(k)]}{(p-k)^2} \nonumber\\
&&p\cos \theta(p)-m\sin \theta(p)=\frac{m_0^2}{4}\int_{\rm PV} \frac{\sin[\theta(p)-\theta(k)]}{(p-k)^2} \ .
\end{eqnarray}
The gap equation has no analytic solution in general. Nevertheless, one can show that the function $\theta(p)$ is odd in $p$, and $\theta(p)\rightarrow \pm \frac{\pi}{2}$ as $p\rightarrow \pm \infty$. One also needs the relation $\tan \theta(p) \rightarrow \frac{p}{m}+{\cal O}(\frac{1}{p})$ as $p\rightarrow \infty$.
In this case, one can diagonalize the Hamiltonian in the large $N_c$ limit in the form

\begin{align}
H=\int \frac{dP}{2\pi}\sum_n E_n(P)m_n^{\dagger}(P)m_n(P)+{\cal O}\left(\frac{1}{N_c}\right) \ ,
\end{align}
where $E_n=\sqrt{P^2+M_n^2}$ is the energy of the meson, and the creation-annihilation operator of the meson state reads

\begin{align}
&m_n(P)=\int \frac{dk}{\sqrt{2\pi N_c|P|}}\times\sum_A \nonumber \\
&\left(b_{A}(P-k)a_A(k)\phi_n^+(k,P)+a_A^{\dagger}(k-P)b^{\dagger}_{A}(-k)\phi_n^-(k,P)\right) \ ,
\end{align}
provided that the wave functions $\phi_n^{\pm}$ satisfy the Bars-Green equation~\cite{Bars:1977ud}

\begin{align}\label{eq:barsgreen}
&\left(E(p)+E(P-p)\mp E_n(p)\right)\phi_n^{\pm}(p,P)=\nonumber \\
&\frac{m_0^2}{2}\int_{\rm PV}\frac{dk}{(p-k)^2}\bigg(C_P(p,k)\phi_n^{\pm}(k,P)-S_P(p,k)\phi_n^{\mp}(k,P)\bigg) \ ,
\end{align}
with

\begin{eqnarray}
C_P(p,k)&=&\cos\bigg(\frac{\theta(p)-\theta(k)}{2}\bigg)\cos\bigg(\frac{\theta(P-p)-\theta(P-k)}{2}\bigg) \nonumber\\
S_P(p,k)&=&\sin\bigg(\frac{\theta(p)-\theta(k)}{2}\bigg)\sin\bigg(\frac{\theta(P-p)-\theta(P-k)}{2}\bigg)\nonumber\\
\end{eqnarray}
The $m_n(P)$ satisfy the commutation relations

\begin{align}
[m_n(P),m^{\dagger}_{n'}(P')]=2\pi\delta_{nn'}\delta(P-P') \ ,
\end{align}
provided that the $\phi_n^{\pm}$ are normalized as

\begin{align}
\int dk \phi_n^{+}(k,P)\phi^{+}_{n'}(k,P)-\phi_n^{-}(k,P)\phi^{-}_{n'}(k,P)=|P|\delta_{n,n'}  \ .
\end{align}
The vacuum state is defined by $m_n(P)|0\rangle=0$, which is  a coherent state in terms of the original quark creation-annihilation operators.  The meson state is given by

\begin{align}
|P,n\rangle=\sqrt{2E_n}m_n^{\dagger}(P)|0\rangle  \ .
\end{align}
The original meson-quark-anti-quark operator can be re-constructed as

\begin{align}
&\frac{1}{\sqrt{N_c}}\sum_A b_{A}(P-k)a_A(k)\nonumber \\
&=\sqrt{\frac{2\pi}{|P|}}\sum_n \bigg(\phi_n^+(k,P)m_n(P)-\phi^-_n(k-P,-P)m_n^{\dagger}(-P)\bigg)
\end{align}
As $P\rightarrow \infty$, the backward moving component $\phi_n^-\rightarrow 0$ at a rate  of $\frac{1}{P^2}$, while the forward component approaches the light-Front wave function $\phi_n^{+}(k,P)\rightarrow \phi_n(\frac{k}{P})$ and is only supported in the interval $0<k<P$. The same light-front wave function can also be obtained from light-cone quantization which will be introduced next.

\subsection{Light-Front Quantization in $A^+=0$ Gauge.}
Here we present an introduction to the light-front quantization of two-dimensional QCD and derive the t$^\prime$Hooft equation for the light-front wave functions. Our convention for light-front coordinates are  $x^{\pm}=\frac 12  (t\pm z)$.  In the light cone gauge $A_-=A^+=0$, the Lagrangian is

\begin{align}
{\cal L}_{1+1}=&\frac 18 (\partial^+A^{a-})^2+
\psi^\dagger_+iD^-\psi_++\psi_-^\dagger i\partial^+\psi_-\nonumber \\ &+im
(\psi_+^\dagger\psi_--\psi_-^\dagger\psi_+)
\end{align}
in the Weyl-like basis with $\gamma^5\psi_\pm=\pm \psi_\pm$.
% and the Dirac matrices
%$\gamma^\mu=(\sigma^2, i\sigma^1)$ and $\gamma^5=\gamma^0\gamma^1=\sigma^3$,
%$\psi^T=(\psi_+, \psi_-)$.
In this gauge $\psi_-=(m/\partial^+)\psi_+$ is a constraint field, and  $A^-$  can be eliminated by
its equation of motion. The corresponding Hamiltonian on the light front with $x^+=0$, follows canonically in the form

\begin{align}
\label{PMINUS}
P^-&=\int \frac{1}{2}(E^aE^a+m\bar \psi \psi) dy^-\nonumber \\
&=-\frac {im^2}2\int dx^-dy^-\psi^\dagger_+(x^-)\,{\theta}(y^--x^-)\,\psi_+(y^-)
\nonumber \\ &-\frac{g^2}4\int dx^-dy^-\psi^\dagger_+T^a\psi_+(x^-)|x^--y^-|\psi^\dagger_+T^a\psi_+(y^-) \ .
\end{align}
The free field reads

\begin{align}
\label{FREE}
\psi_{+A}(z^-)=\int_0^\infty\frac{dk^+}{2\pi}\left(e^{-ik^+z^-}a_A(k^+)+e^{+ik^+z^-}b_A^\dagger(k^+)\right)
\end{align}
 and the  anticommuting rules for the creation-annihilation operators are

\be
\label{COM}
\left[a_A(k^+),a_B^\dagger(p^+)\right]_+&=\left[b_A(k^+),b_B^\dagger(p^+)\right]_+\nonumber \\
&=\delta_{AB}(2\pi)\delta(k^+-p^+) \ .
\ee
The bound states associated to (\ref{PMINUS}) are eigenstates to $P^-\left|P,n\right>=P_n^-\left|P,n\right>$,
subject to Gauss law (color singlet).
% Here, the label-n is shown explicitly in reference to the   n-th meson bound state.
In the large $N_c$ limit and weak coupling with $m\gg m_0$ pair production is suppressed and mesons
and baryons are stable. In leading order in $1/N_c$, a meson with a pair of $q\bar q$ is represented by the colorless state

\begin{align}
\label{MESONWF}
\left|P,n\right>=
\int_0^{1}\frac{P^+dx}{{\sqrt{\pi}}}\frac{\varphi_n(x)}{\sqrt{N_c}}a^\dagger_A(xP^+)b^\dagger_A((1-x)P^+)\left|0\right>
\end{align}
with Feynman $x=k^+/P^+$ such that $0\leq x\leq 1$ and $x+\bar x=1$.

In this  construction, it is not difficult to show that the amplitudes $\varphi_n(x)$ solve the $^\prime$t Hooft equation~\cite{tHooft:1974pnl}

\begin{align}
\label{THOOFT}
%\bigg(\frac {m^2}{x}+\frac {m^2}{\bar x}\bigg)
\frac {m^2}{x\bar x}\varphi_n(x)-m_0^2\int_{\rm PV} dy\frac{\varphi_n(y)-\varphi_n(x)}{(x-y)^2}=M_n^2\varphi_n(x) \ ,
\end{align}
with the canonical mass scale $m_0^2=\lambda/\pi$ and $\lambda=g^2N_c$ the t$^\prime$Hooft coupling. If  we take  the large $P$ limit in the Bars-Green equation  Eq.~(\ref{eq:barsgreen}), then we obtain the same $^\prime$t Hooft equation here for $\phi_n^{+}(k)\rightarrow \varphi_n(x)$ where $x=\frac{k}{P}$. Thus, by taking the large momentum limit of the equal-time wave function we  recover the light-Front wave function.

\section{Hadron Mass/Energy sum rule in Equal-time quantization}\label{MASSS}
To study the mass structure of a hadron, one can start from the energy momentum tensor
of the theory~\cite{Ji:1994av}. Massive spinor QCD is characterized by the symmetric energy momentum tensor
\bea
\label{1}
T^{\mu\nu}=&&\frac 2{\sqrt{-g}}\frac{\delta S_{1+1}}{\delta{g_{\mu\nu}}}\nonumber\\
=&&F^{a\mu\lambda}F^{a\nu}_\lambda+\frac 14 g^{\mu\nu}F^2+\frac 12 \overline\psi \gamma^{[\mu} i\overleftrightarrow D^{\nu]_+}\psi
\eea
with $\overleftrightarrow{D}=\overrightarrow{D}-\overleftarrow{D}$ and $[]_+$ denotes symmetrization. In terms of   $E^a$, the stress tensor reads

\bea
\label{3}
T^{\mu\nu}=\frac 12 g^{\mu\nu}E^aE^a+\frac 12 \overline\psi \gamma^{[\mu} i\overleftrightarrow D^{\nu]_+}\psi
%-\frac 12 \overline\psi \gamma^{[\mu} i\overleftrightarrow D^{\nu]_+}\psi +mg^{\mu\nu} \overline\psi\psi\nonumber\\
\eea
It is conserved $\partial_\mu T^{\mu\nu}=0$, with a non-vanishing trace

\be
\label{3X}
T^\mu_\mu=E^aE^a+m\overline\psi\psi
\ee
%after using the fermionic equation of motion $(i\slashed D-m)\psi=0$.
In two-dimensions,
QCD is super-renormalizable, hence non-conformal. Amusingly, the trace part in
(\ref{3X}) resembles the trace part of QCD with a dimensionless coupling in four dimensions~\cite{Ji:1994av}.

\subsection{Virial Theorem}

The energy-momentum tensor can be decomposed as the sum of a traceless and trace part~\cite{Ji:1994av}

\begin{align}
T^{\mu\nu}=\hat T^{\mu\nu}+\frac  12 {g^{\mu\nu}}T^\alpha_\alpha \ ,
\end{align}
where the traceless part reads

\begin{align}
\hat T^{\mu\nu}=\frac 12 \overline\psi \gamma^{[\mu} i\overleftrightarrow D^{\nu]_+}\psi-\frac{g^{\mu\nu}}{2}m\bar\psi \psi ,
\end{align}
and the trace part is given in Eq.~(\ref{3X}).

For a single-particle state $|P,n\rangle$ with the standard normalization $\langle P,n|P',n'\rangle=2E_n\delta_{n,n'}(2\pi)\delta(P-P')$, one has

\begin{align}\label{fowardT}
\langle P,n|T^{\mu\nu}|P,n\rangle=2P^{\mu}P^{\nu} \ .
\end{align}
By considering the ${00}$ component in the rest frame, one finds the relation between the matrix elements

\begin{align}\label{T00}
&\langle 0,n| \bar \psi (-i\gamma^zD_z)\psi|0,n\rangle+\langle 0,n|m\bar \psi\psi|0,n\rangle\nonumber \\
&+ \langle 0,n|\frac{1}{2}E^aE^a|0,n\rangle=2M_n^2 \ .
\end{align}
By taking the trace, one has in any frame

\begin{align}\label{Trace}
\langle P,n|T^\mu_\mu|P,n\rangle=\langle P,n|m\bar \psi\psi+E^aE^a|P,n\rangle=2M_n^2 \ .
\end{align}
which basically sets the scales for the theory.

With this in mind, lets us consider the mass sum rule in the rest frame. The Hamiltonian can be obtained from the $T^{00}$ component

\begin{align}
H&=\hat H+\bar H \nonumber \\
&=\int dz \bigg(\bar \psi i\vec{\gamma}\cdot\vec{D}\psi +m\bar \psi\psi+\frac{1}{2}E^aE^a\bigg) \ ,
\end{align}
where the traceless part $\hat H$ and the trace part $\bar H$ read

\begin{eqnarray}
\hat H&=\int dz \hat T^{00}=\int dz \bar \psi \bigg(-i\gamma^zD_z +\frac{1}{2}m\bigg)\psi \ ,\nonumber\\
\bar H&=\int dz \frac{1}{2}T^{\mu}_{\mu}=\int dz (\frac{1}{2}E^aE^a+\frac{1}{2}m\bar\psi\psi) \ .
\end{eqnarray}
If we define the average

$$\langle O \rangle_{P,n}=\frac{\langle P,n|O|P,n\rangle}{\langle P,n|P,n\rangle}$$
and notice that $\int dz= L=2\pi \delta_p(0)$, we can read from  Eq.~(\ref{Trace}) that the contribution of the trace part to the total mass in the rest frame is half of the mass, which is the relativistic Virial theorem first discussed in Ref.~\cite{Ji:1994av}

\begin{eqnarray}
\langle\bar H\rangle_{0,n}&=&\frac{M_n}{2}  \ ,\nonumber \\
\langle\hat H\rangle_{0,n}&=&\frac{M_n}{2}   \ .
\end{eqnarray}
More explicitly, one obtains the mass-sum rule

\begin{align}\label{MASS}
&\langle\bar H\rangle_{0,n}={\cal G}+ \frac{1}{2}{\cal M}=\frac{M_n}{2}\nonumber \\
&\langle\hat H\rangle_{0,n}={\cal K}+\frac{1}{2}{\cal M}=\frac{M_n}{2}  \ .
\end{align}
or more succinctly,

\begin{equation}
  {\cal K}+ {\cal G} + {\cal M} = M_n
\end{equation}
with

\begin{align}\label{eq:defMKG}
&{\cal M}=\langle \int dz m\bar \psi\psi\rangle_{0,n}=\frac{\langle 0,n|m\bar\psi \psi|0,n\rangle}{2M_n} \ ,\nonumber \\
&{\cal K}=\langle\int dz\bar \psi (-i\gamma^zD_z)\psi\rangle_{0,n}=\frac{\langle 0,n|\bar \psi (-i\gamma^zD_z)\psi|0,n\rangle}{2M_n} \ , \nonumber\\
&{\cal G}=\langle\int dz \frac{1}{2}E^aE^a\rangle_{0,n}=\frac{\langle 0,n|\frac{1}{2}E^aE^a|0,n\rangle}{2M_n} \ .
\end{align}
Comparing the two equations in Eq.~(\ref{MASS}), one obtains in the rest frame

\begin{align}\label{eq:K=G}
{\cal K}={\cal G} \ ,
\end{align}
The fermion kinetic and gluon energies in the rest frame are equal. This is consistent with the Virial theorem $\langle |p|\rangle=\langle r\frac{\partial V}{\partial r}\rangle$ for a  relativistic quantum mechanical system~\cite{Lucha:1989jf},  with the hamiltonian $H=|p|+V(r)$ and a linear potential $V(r)=\sigma r$, if the gluon energy is identified with the average of the linear potential energy. This is also consistent with the results in Ref.~\cite{Jia:2018mqi} for massive state $\bar ss$, $\bar cc$ and in the $m\rightarrow \infty$ limit. This can also be obtained from a Feynman-Hellman theorem argument which is given in Appendix~\ref{virial}.

\subsection{Partonic contributions to ${\cal G, K, M}$}

We now evaluate  explicitly the matrix-elements above. We start with the scalar matrix element $\langle P,n|m\bar\psi \psi|P,n\rangle$. It can be calculated in a
generic frame in terms of the wave-functions $\phi_n^{\pm}$ in Eq.~(\ref{eq:barsgreen}) as

\begin{align}
&\langle P,n|m\bar\psi \psi|P,n\rangle=\frac{2mE_n(P)}{|P|}\nonumber \\
&\times \int dp\left[\cos \theta(p)+\cos \theta(\bar p) \right]\left[|\phi_n^+|^2+|\phi_n^-|^2\right](k,P) \ .
\end{align}
Since the matrix element is independent of $P$, by taking the $P\rightarrow \infty$ limit one obtains in terms of the light-front wave function in Eq.~(\ref{THOOFT})

\begin{align}
\langle P,n|m\bar \psi \psi|P,n\rangle=2m^2\int_{0}^1 dx\frac{\varphi_n^{\dagger}(x)\varphi_n(x)}{x\bar x} \ .
\end{align}
Therefore
\begin{align}
{\cal M}=\frac{m^2}{M_n}\int_{0}^1 dx\frac{\varphi_n^{\dagger}(x)\varphi_n(x)}{x\bar x} \ ,
\end{align}
gives the quark mass contribution.

Similarly, for the gluonic contribution one obtains in term of the light-front wave function

\begin{align}\label{GXX}
&\langle P,n|\frac{1}{2}E^aE^a|P,n\rangle\nonumber \\
&=-m_0^2{\rm PV}\int_{0}^1 dxdy\frac{\varphi_n^{\dagger}(x)\left(\varphi_n(y)-\varphi_n(x)\right)}{(x-y)^2} \ ,
\end{align}
and

\begin{align}
&{\cal G}=-\frac{m_0^2}{2M_n}{\rm PV}\int_{0}^1 dxdy\frac{\varphi_n^{\dagger}(x)\left(\varphi_n(y)-\varphi_n(x)\right)}{(x-y)^2}  \ .
\end{align}
which is again frame-independent.

The kinetic energy contribution in the rest frame can be calculated as

\begin{align}\label{eq:genericK}
&\langle P,n|\bar\psi(-i\gamma^zD_z) \psi|P,n\rangle=\frac{2E_n(P)}{|P|}\nonumber \\
&\times \int dp\left[p\sin \theta(p)+\bar p\sin \theta(\bar p) \right]\left[|\phi_n^+|^2+|\phi_n^-|^2\right](p,P) \ .
\end{align}
By using the virial theorem in Eq.~(\ref{eq:K=G}), one can show that the kinetic energy in the rest frame is equal to
\begin{align}\label{eq:restK}
{\cal K}=-\frac{m_0^2}{2M_n}{\rm PV}\int_{0}^1 dxdy\frac{\varphi_n^{\dagger}(x)\left(\varphi_n(y)-\varphi_n(x)\right)}{(x-y)^2} \ .
\end{align}
It is non-trivial to prove this using the Bars-Green equation directly. However,   using the bosonized Hamiltonian in the  large $N_c$ limit,
it can be shown directly  that ${\cal G}={\cal K}$. Using the explicit form of the t'Hooft equation, it is easy to see that Eq.~(\ref{MASS}) is satisfied.

To summarize, in the rest frame, the mass of a hadron in  two dimensional  QCD comes from
three sources:  the quark kinetic energy ${\cal K}$, the gluon potential energy ${\cal G}$, and finally the quark mass ${\cal M}$.
The first two are equal with ${\cal K=G}$. In Sec.~\ref{GRAVITON}, we show how the gluonic content of the hadron mass can be related to experimentally measurable quantities.

\subsection{Relation to an observable}

To relates the mass sum rule to an observable, one may consider
the contribution from the traceless part of the energy-momentum tensor,
which is a twist-two operator. The light-front sum rule,
can relate the $T^{++}$ component of the stress tensor to the twist-two parton distribution
function.  As  shown in Ref.~\cite{Jia:2018mqi}, $\frac{1}{2}T^{++}$ dominates the traceless part $\hat T^{00}$ of $T^{00}$ in the large momentum frame

\begin{align}
\langle P,n|\bar \psi\left(-i\gamma^zD_z+\frac{m}{2}\right)\psi|P,n\rangle \rightarrow (P^+)^2+{\cal O}(\frac{1}{P^+}) \ .
\end{align}
In  two dimensions, there is only one quark twist-two operator and no gluon twist-two operator as there is
no dynamical gluon which can carry the hadron momentum. One can define the quark parton distribution function(PDF) for the meson state

\begin{align}
q_n(x)=\int \frac{dx^-}{4\pi}e^{-ix^-P^+x}\langle P,n|\bar \psi(x^-)\gamma^+[x^+,0]\psi(0)|P,n\rangle \ .
\end{align}
It can be calculated straightforwardly as
\begin{align}
q_n(x)=|\varphi_n(x)|^2 \ .
\end{align}
The first moment of the quark-PDF is related to the traceless part of $T^{\mu \nu}$ through

\begin{align}
\langle P,n| T^{++}|P,n\rangle=4(P^+)^2\int_0^1 dx x q_n(x) \ .
\end{align}
This is consistent with the normalization

$$\langle P,n| T^{++}|P,n\rangle=2(P^+)^2\,,$$
since one has

\begin{align}
\int_0^1 dx x q_n(x) =\frac{1}{2}\int_0^1(x+1-x)|\varphi_n(x)|^2=\frac{1}{2}
\end{align}
due to the relation $\varphi_n(1-x)=(-1)^n\varphi_n(x)$. Therefore,  the
twist-two light-front sum rule does not provide an additional information other than
that the kinetic energy ${\cal K}$ of the quark plus  half the quark mass term $\frac 12{\cal M}$ contribute half of the meson mass.

\section{IMF Sum Rule and Light-Front Distribution Functions}\label{DISTRIBUTION}

Let us consider the mass-sum rule in the IMF. The light-cone Hamiltonian is related to the $T^{+-}$ component of the stress tensor

\begin{align}
T^{+-}&=\frac{1}{2}E^aE^a+\frac{1}{2}\bar \psi i\gamma\cdot D\psi \nonumber \\
&=\frac{1}{2}E^aE^a+\frac{1}{2}m\bar \psi \psi \ ,
\end{align}
where in the second line, the equation of motion has been used. The mass sum-rule
in the light front then reads

\begin{align}
P^-=\frac{M_n^2}{2P^+}=\langle\int dx^-\frac{1}{2}m\bar \psi \psi\rangle_{P,n}+\langle\int dx^-\frac{1}{2}E^aE^a\rangle_{P,n} \ .
\end{align}
which is

\begin{equation}
\label{SRIMF}
    2{\cal G} + {\cal M} = M_n,
\end{equation}
by using Eq.~(\ref{eq:defMKG}) and the normalization $\langle P|P\rangle=P^+L^-$ where $L^-=\int dx^-$ is the length along the light-cone spatial direction $x^-$. Here both terms are higher-twist~\cite{Ji:2020baz} where the twist are defined as the mass dimension minus the longitudinal spin. This is consistent with the light-cone sum-rule in Ref.~\cite{Jia:2018mqi}. We now discuss how the light-cone mass sum-rule can be related to moments of twist-three and twist-Four gluon and quark distributions functions.
This will help understand the possible zero-mode contributions at $x=0$.

\subsection{Fermionic and gluonic distributions}

The fermionic mass contribution to the mass sum rule is directly probed by the chiral-odd scalar quark bi-local
(twist-three light-front distribution)~\cite{Ji:2020baz} \be
\label{QX}
Q_n(x)=\frac{P_n^+}{2M^2_n}\int \frac{dz}{2\pi}e^{ixP_n^+z}\left<P,n\left|m\overline\psi(0)[0,z]\psi(z)\right|P,n\right>\nonumber\\
\ee
with the zeroth  moment fixed

\be
\label{QZERO}
Q_{n,0}=\int dx \,Q(x)=\frac{\left<P,n\left|m\overline\psi\psi\right| P,n\right>}{2M^2_n}
\ee
The gluon mass contribution can be probed by the twist-four
gluon gauge-invariant bi-local

\begin{align}
\label{FFX}
F_n(x)&=\frac{P_n^+}{4M_n^2}\int \frac{dz}{2\pi}e^{ixP_n^+z}\left<P,n\left|F_{\mu\nu}(0)[0,z]F^{\mu\nu}(z)\right|P_n\right> \nonumber \\
&=\frac{P_n^+}{2M^2_n}\int \frac{dz}{2\pi}e^{ixP_n^+z}\left<P,n\left|E(0)[0,z]E(z)\right|P,n\right> \ ,
\end{align}
with the zeroth moment

\be
\label{ZEROF}
F_{n,0}=\int dx \,F_n(x)=\frac{\left<P,n\left|E^aE^a\right| P,n\right>}{2M_n^2}
\ee
The light-front mass-sum rule then reads

\begin{align}
Q_{n,0}+F_{n,0}=1 \ .
\end{align}
Below we discuss each of these two distribution functions in detail, particularly
the possible existence of a distribution $\delta (x)$ which may ruin the practical sum rule.

\subsection{Chiral-odd quark parton distribution and light-cone sum rule}\label{QUARK}
In light-front quantization, zero modes are important. In fact,
the chiral condensate in the vacuum comes entirely from zero modes.
It has been suggested that $Q_n(x)$ has a regular and singular part~\cite{Efremov:2002qh}

\be
Q_n(x)=\delta(x)\,Q_{n,0}+Q_{n,\rm reg}(x)
\ee
provided that the zeroth moment of the regular part vanishes.
The delta-function maybe viewed as a signal of the quark condensate on the light cone.
%Its occurence at $x=0$ makes it hard to access experimentally. This notwithstanding,
Its possible  presence affects the  partonic  sum rules, a point of recent  emphasis~\cite{Aslan:2018tff,Ji:2020baz,Bhattacharya:2020jfj}.

\subsubsection{Analysis in light cone gauge}

Eq. (\ref{QX}) can be readily calculated in the light cone gauge using the canonical formulation
briefly reviewed before. More specifically, in the light cone gauge

\begin{widetext}
\be
\label{QLINE}
\overline\psi(0)[0,z]\psi(z)=\psi_+^\dagger(0)(-i\psi_-)(z)+(-i\psi_-)^\dagger(0)\psi_+(z)=
\psi_+^\dagger(0)(-i(m/\partial^+)\psi_+)(z)+(-i(m/\partial^+)\psi_+)^\dagger(0)\psi_+(z)
\ee
\end{widetext}
The fermionic field in (\ref{QLINE}) is  given in (\ref{FREE}).
The fact that the expectation value in (\ref{QLINE}) is multiplied by $m$ is natural, since the chiral odd operator $\overline\psi\psi$
creates a left-right pair which in the light cone and $m=0$ cannot interact. It is however uncorrect to conclude
that the matrix element vanishes as $m\rightarrow 0$ as we now show.

Inserting (\ref{QLINE}) into (\ref{QX}), and using the bound meson
wavefunction (\ref{MESONWF}) in leading order in $1/N_c$,  lead to a string of contractions in the vacuum.
A repeated use of (\ref{COM}) yields the result

\be
\label{QNX}
Q_n(x)=\frac{2m}{ M_n}\frac{|\varphi_n(x)|^2}x
\ee
In agreement with the Weisberger relation for the parton model in four dimensions~\cite{Weisberger:1972hk,Brodsky:2007fr}.
We  have made explicit the index-$n$ for an $n$-bound meson state. The occurence of the overall factor of $m$
reflects on the chiral odd character of the bi-local operator.
The zeroth moment of (\ref{QNX}) is  dominated by the singular behavior of the parton distribution funtion
near the edge in (\ref{EDGE}), i.e. $\varphi_n(x)\sim C_n(x\bar x)^\beta$ as reviewed in Appendix~\ref{LIGHT}. The result is finite

\be
Q_{n,0}=\frac{2m}{ M_n}\int_0^1 dx\,\frac{|\varphi_n(x)|^2}x=\frac{C^2_n\pi m_0}{M_n\sqrt{3}}
\ee
even in the chiral limit.
A comparison with (\ref{QZERO}) implies the scalar condensate in the $n$-bound state
\be
\label{SIGMA}
\left<P,n\left|\overline\psi\psi\right| P,n\right>=C^2_n\frac{2\pi m_0}{\sqrt{3}}=-4\pi C^2_n\frac{\left<\overline\psi\psi\right>}{N_c}
\ee
The last identity makes use of (\ref{QQBAR}) below.

While the finite sum rule (\ref{QZERO}) is fulfilled  for all bound states, the chiral odd quark distribution (\ref{QNX}) {does not develop} a  delta-function! Rather,
the rapid vanishing of the wavefunction or parton distribution amplitude (PDA) at the edge with the exponent $\beta\sim m$ is what enforces the sum rule. In retrospect, this is expected since
the chiral odd character of the operator generates a pre-factor $m$ that requires compensation for a finite zeroth moment. In a way, the vacuum physics is
encoded in the $x\sim 0$ region of the chiral odd parton distribution function, {not in the way it diverges but in the way it vanishes}

\bea
&&Q_n(x\sim 0)\sim \frac{2m}{M_n} C_n x^{2\beta-1}\nonumber\\
&&2\beta-1=\frac{2\sqrt 3}{\pi}\,\frac m{m_0}-1\gg 0
\eea
where the inequality is in the weak coupling limit~\cite{Zhitnitsky:1985um}.
The chiral limit  and the $x\rightarrow 0,1$ limits are subtle. These observations are
general,  and may carry to QCD in  four dimensions, although in  the latter instanton effects are a strong source of chirality flips without the current mass $m$ penalty~\cite{Shuryak:2020ktq}.

\subsubsection{The special case with $n=0$: \\the would-be Goldstone mode}
In two-dimensions and finite $N_c$ there is no spontaneous breaking of chiral symmetry owing to Coleman theorem~\cite{Coleman:1973ci}.
At large $N_c$ and weak coupling,  chiral symmetry is almost broken due to the BKT mechanism with the appearance
of an almost Golstone mode~\cite{Witten:1978qu}, the analogue of the pion and a finite chiral condensate~\cite{Zhitnitsky:1985um,Burkardt:1994xu}

\be
\label{QQBAR}
\left<\overline\psi \psi\right>=-\frac{N_cm_0}{\sqrt{12}}
\ee
For the would-be Goldstone or pion mode with  $n=0$, $C_0=1$~\cite{Zhitnitsky:1985um}  and (\ref{SIGMA}) appears to obey a  chiral reduction rule in the massless limit.
Indeed, since $\overline\psi\psi$ is a scalar its value in a moving state is the same as its value  in the state at rest. Therefore,

\be
\label{Q5}
\left<P,0\left|\overline\psi\psi\right| P,0\right>=-2\left<\bigg[\frac {Q_5}{f_0}, \bigg[\frac {Q_5}{f_0}, \overline\psi\psi\bigg]\bigg]\right>=-2\frac{\left<\overline \psi\psi\right>}{f_0^2}\nonumber\\
\ee
where  $Q_5$ is the axial U(1) charge operator,
and $f_0=\sqrt{N_c/2\pi}$ is the analogue of the pseudoscalar decay constant. This identification is consistent with (\ref{DECAY})  for $n=0$ since  by chiral reduction we can check that
at rest

\begin{align}\label{DECAY0}
\left<0\left|\overline\psi i\gamma^5\psi\right|P,0\right>=-\frac {\sqrt 2}{f_0}\left<\overline \psi\psi\right>
=\sqrt{2}f_0\bigg[\frac m2\int_0^1dx \frac{\varphi_0(x)}{x\bar x}\bigg]
\end{align}
In particular, the vacuum chiral condensate is directly tied to the would-be Goldstone mode light
cone wavefunction   through

\be
\label{MPSI}
\left<\overline\psi\psi\right>=-\frac 12 f_0^2m\int_0^1 dx\frac{\varphi_0(x)}{x\bar x}
\ee
 %The last identity provides for a {\bf definition} of the chiral condensate in terms of  the light cone wavefunction for the would-be Goldstone mode.

We now note that  the Feynman-Hellman theorem for this mode gives

\be
\frac{\partial M_0^2}{\partial m}=\left<P,0\left|\overline\psi\psi\right| P,0\right>
\ee
which together with (\ref{Q5})  imply the Gell-Mann-Oakes-Renner (GOR) like relation

\be
f_0^2M_0^2=-2 m\left<\overline\psi\psi\right>
\ee
It follows that 
 the gluon contribution in the tracefull part of the mass in (\ref{Trace}) 
is equal to the chiral symmetry breaking contribution,

\be
\left<P,0|E^aE^a|P,0\right>=\left<P,0|m\overline\psi\psi|P,0\right>=M_0^2
\ee
This result is in agreement with an earlier observation made by one of us in 4-dimensions~\cite{Ji:1995sv}.

\subsection{Twist-four gluon parton distribution and light-cone sum rule}\label{GLUON}
Besides the quark contribution, the gluon part could also contains a regular and singular contribution~\cite{Ji:2020baz}

\be
F_n(x)=\delta(x)\,F_{n,0}+F_{n,\rm reg}(x)
\ee
assuming that the zeroth moment of the regular part vanishes.
The occurrence of the delta-function maybe viewed as a signal of the gluon condensate on the light-front,
with the wee-gluons at $x=0$ collecting into a superfluid-like component, in an
otherwise normal component of a hadron~\cite{Ji:2020baz}.  This point was recently
discussed in the context of perturbation theory in~\cite{Hatta:2020iin}.
We now proceed to evaluate  (\ref{FFX}) non-perturbatively in two dimensions.

In the light cone gauge with normals $n^{\mu}_\pm=(1, \pm 1)$,

\be
F^{+-a}=\partial^+A^{-a}=n_+^\mu n_-^\nu F_{\mu\nu}^a=2E^a
\ee
the equation of motion for the gauge field  simplifies to

\be
\label{DPE}
\partial^+E^a={2g}\psi_+^\dagger T^a\psi_+
\ee
Inserting (\ref{DPE}) into (\ref{FFX}) yields

\begin{widetext}
\bea
\label{FFXX}
F_n(x)=&&\frac{P_n^+}{2M_n^2}\int \frac{dz}{2\pi}e^{ixP_n^+z}\left<P,n\left|\bigg(\frac{2g}{\partial^+}\psi_+^\dagger T^a\psi_+(0)\bigg)\bigg(\frac{2g}{\partial^+}\psi_+^\dagger T^a\psi_+(z)\bigg)\right|P,n\right>\nonumber\\
=&&\frac{P_n^+}{2M_n^2}\int \frac{dz}{2\pi}e^{ixP_n^+z}\left<P,n\left|\bigg(2g\int_0^\infty dy\, \psi_+^\dagger T^a\psi_+(y)\bigg)\bigg(2g\int_z^\infty dw\, \psi_+^\dagger T^a\psi_+(w)\bigg)\right|P,n\right>
\eea
\end{widetext}
Substituting the free field decomposition (\ref{FREE}) into (\ref{FFXX})  yield 1-body (self-energy) and 2-body type contractions (cross terms).
The final result is
\begin{widetext}

\bea
\label{FINALGLUE}
xF_n(x)=&&\frac{m_0^2}{4M_n^2x}\bigg(\int^{1}_{{\rm max}(0,x)}dx_1\,\left(|\varphi_n(x_1)|^2+|\varphi_n(\bar x_1)|^2\right)\nonumber\\
&&-\int^{{\rm min}(1,1+x)}_{{\rm max}(0,x)}dx_1\,\left(\varphi_n(x_1)\varphi_n^{\dagger}(x_1-x)+\varphi_n(\bar x_1)\varphi_n^{\dagger}(\bar x_1+x)\right)\bigg) \ .
\eea
\end{widetext}
As $x\rightarrow 0$, the self-energy and cross terms cancel with each other due to color neutrality, leading
to a finite gluon distribution $xF_n(x)$ at $x=0$,  a characteristic of color neutral bound-states.  This cancellation is somehow
reminiscent of the cancellation of the  gauge dependence between the vertex and self-energy in the derivation of the
$^\prime$t Hooft equation~\cite{tHooft:1974pnl}.
The zeroth moment (\ref{ZEROF}) reproduces the gluon-condensate~(\ref{GXX}) in agreement with the t$^\prime$Hooft equation (\ref{THOOFT}) and the mass sum rule (\ref{MASS}) in the chiral limit. Notice that (\ref{GXX}) vanishes for the pion state $\varphi_0(x)\sim \theta(x\bar x)$ as it should.
No singular distribution is seen to arise in (\ref{FINALGLUE}), yet the relation (\ref{ZEROF}) is satisfied.

%\begin{widetext}

\section{Form factors of the energy-momentum tensor}\label{GRAVITON}

Apart from the light-cone sum rules, an empirical  way to measure the gluon contribution to
the mass sum rule of the nucleon, is through
diffractive photo- or electro-production of heavy mesons off  nucleons near treshold as suggested intially in~\cite{Kharzeev:1998bz},
and pursued experimentally in~\cite{Hafidi:2017bsg,Ali:2019lzf}. The threshold cross section is dominated by glueball-like
exchanges with tensor and scalar quantum numbers. A detailed analysis shows that the exchange probes the nucleon
gravitational form factor with the glueball as a spin-2 exchange, allowing for a partial extraction of the gluon contribution to the nucleon mass~\cite{Hatta:2018ina,Mamo:2019mka}. The importance
of the role of the gluons in the nucleon mass was emphasized earlier by one of us~\cite{Ji:1994av}.
%To understand further the nonperturbative role of the gluons  and quarks in the composition of the gravitational form factor, we propose to analyze it in two-dimensions.

The gravitational form factor in a two-dimensional meson state is contrained by Lorentz symmetry, parity  and energy-momentum conservation.
 Its general decomposition under these strictures leads the  invariant form factors
\begin{align}
\label{THETA123}
\left<p_2\left|T^{\mu\nu}\right|p_1\right>=\frac 12 (g^{\mu\nu}q^2-{q^\mu q^\nu})\tilde \Theta_1(q^2)
+\frac 12 {p^\mu p^\nu}\tilde \Theta_2(q^2)
\end{align}
with $q^\mu=p^\mu_2-p_1^\mu$ and $p^\mu=p_1^\mu+p_2^\mu$. Note that the transverse axial-vector $\tilde{q}^\mu=\epsilon^{\mu\lambda}q_\lambda$
does not introduce an independent 2-tensor since $\tilde{q}^\mu\tilde{q}^\nu=q^\mu q^\nu-g^{\mu\nu}q^2$.
In  4-dimensions, the two form factors correspond to the spin representations
$1\otimes 1=0\oplus 1\oplus 2$ with  $1$ excluded by parity. They reflect on the tensor exchange or graviton ($2$), and
the scalar exchange or dilaton (0).

We now note that in 2-dimensions

\begin{align}
{p^\mu p^\nu}=\frac{4M^2-q^2}{q^2}(g^{\mu\nu}q^2-{q^\mu q^\nu}) \ ,
\end{align}
for on-shell momenta $p_{1,2}$. Thus, the two form factors $\tilde\Theta_{1,2}$ can always be combined
to a single form factor

\begin{align}
\label{THETA124}
\left<p_2\left|T^{\mu\nu}\right|p_1\right>=\frac 12 {p^\mu p^\nu}\Theta_2(q^2) \ .
\end{align}
which is consistent with the $m_1\otimes m_2=m_1+m_2$ for the irreducible representation of U(1).  The dilaton form factor is described by the trace

\be
\label{TT}
%\Theta_2(q^2)=\frac {\left<p_2\left|2T^\mu_\mu\right|p_1\right>}{4M^2-q^2}
\left<p_2\left| T^\mu_\mu\right|p_1\right>=\frac 12(4M^2-q^2)\Theta_2(q^2)
\ee
The squared meson masses $M^2$ are fixed by the transcendental $^\prime$t Hooft equation introduced before.
The normalization $\Theta_2(0)=1$ is fixed by recalling that $H=\int dx\, T^{00}$ is the Hamiltonian, with
$\left<p_1\left|H\right|p_1\right>=p_1^0[2p_1^0(2\pi)\delta_p(0)]$.

%The invariant form factors $\Theta_{1,2}$ are fixed by the energy density
%$T^{00}$ and the trace identity (\ref{TT})

%\begin{align}
%2\bigg(M^2+\frac{Q^2}4\bigg)\Theta_2(Q^2)=\left<p_2|T^\mu_\mu|p_1\right>
%\end{align}
%in the Breit frame  with  $q^\mu=(0,Q)$ and $p^\mu_{1,2}=(E_p,\mp Q/2)$.

The dilaton  form factor is mediated by the exchange of a massive  state of mass $2M$  when recast in the form

\be
\Theta_2(q)=\frac {\left<p_2\left|2T^\mu_\mu\right|p_1\right>}{4M^2-q^2}
\ee
%with the numerator expanded  using (\ref{MASS})  and (\ref{eq:K=G}), in agreement with the normalization
%$\Theta_2(0)=1$.
providing for a direct measure of the  gluon content of the meson state

\begin{align}
 \label{FF12}
\Theta_2(q^2)=\frac 2{4M^2-q^2}\bigg<p_2\bigg|\bigg(E^aE^a+m\overline\psi\psi\bigg)\bigg|p_1\bigg>
\end{align}
in the chiral limit.
Near the production treshold of the mesons, the dilaton  probes the trace part
of the energy-momentum tensor in a bound hadron in two-dimensional QCD.  We interpret the $2M$ state as a massive $^{\prime\prime}$longitudinal-glueball$^{\prime\prime}$ resolving into
a two-meson state by mixing, for this point see also Eq.~(\ref{10X}) and discussion below. This observation maybe useful for four dimensional QCD, where the trace
part probes the trace anomaly say in a bound nucleon state, as pursued experimentally  in
diffractive photo-production of heavy mesons off nucleon targets~\cite{Hafidi:2017bsg,Ali:2019lzf}. The diffractive production involves
the exchange of bulk gravitons  as dual of boundary massive tensor and scalar glueballs in  holographic QCD~\cite{Mamo:2019mka}.

\section{Dual mass sum rule}~\label{ANOMALYX}
Finally, we now show that the static electric field energy is related to the Coulomb energy represented by chirally symmetric
4-Fermi  interactions.  This reasoning will show how the non-perturbative aspects of the vacuum  in the equal-time form are tied to a fermionic
mass sum rule for  the bound mesons, that is dual to the gluonic mass sum rule in the instant form.

We recall that the colored axial-vector current which is the difference  between the gauge-invariant currents for left and right-handed fermions is anomalous

\be
\label{7}
D^{ab}_\mu\bigg(\overline\psi \gamma^\mu\,\gamma^5\,T^b\psi\bigg)=-\frac {g_0}{4\pi}\epsilon_{\mu\nu}F^{a\mu\nu}=-\frac{g_0}{2\pi}E^a
\ee
as can be checked  from the one-loop vacuum polarization diagram.
Using (\ref{2}) and the equation of motion for the gauge fields $
D^{ab}_\mu F^{b\mu\nu}=-g\overline\psi \gamma^\nu\,T^a\psi $,
we have

\be
\label{8}
\epsilon^{\mu\nu}D^{ab}_\mu E^b=-g_0\overline\psi \gamma^\nu\,T^a\psi=-g_0\epsilon^{\nu\mu}\overline\psi \gamma_\mu\,\gamma^5\,T^a\psi
\ee
The last equality follows from the fact that the vector and axial-vector currents are dual in two-dimensions,
$\overline{\psi} \gamma^\nu\,T^a\psi=-\epsilon^{\nu\mu}\overline\psi \gamma_\mu\,\gamma^5\,T^a\psi $.
Therefore we have

\be
\label{9}
D^{ab}_\mu E^b=g_0\overline\psi \gamma_\mu\,\gamma^5\,T^a\psi
\ee
which yields the exact covariant screening equation for the dual potential~\cite{Belvedere:1978fj}

\be
\label{9X}
\bigg(D^2+\frac{m_0^2}{2N_c}\bigg)^{ab}E^b=0
\ee
with the canonical mass scale $m_0^2=g_0^2N_c/\pi$.
%and $\lambda=g_0^2N_c$ the t$^\prime$Hooft coupling.
The emergent screening mass squared  $m_0^2/2N_c$ due the  light quarks is subleading in $1/N_c$ as it should. It will
be important to keep it as we now show.

We now note the gauge invariant identity

\begin{widetext}
\be
\label{10XX}
\partial^2(E^aE^a)=2(\partial_\mu E^a)(\partial^\mu E^a)+2(\partial^2E^a)E^a\equiv
2(D_\mu^{ab}E^b)(D_\mu^{ac}E^c)+2((D^2)^{ab}E^b)E^a
\ee
where in the last equality manifest gauge invariance is enforced
since the starting  expression
$\partial^2(E^aE^a)$
is gauge invariant. Using (\ref{9}) and (\ref{9X}) in (\ref{10XX})  we obtain

\bea
\label{10}
\bigg(\partial^2+\frac {m_0^2}{N_c}\bigg) (E^aE^a)
=
%=2(D^2E^a)E^a+2(D_\mu E^a)D^\mu E^a)\nonumber\\
2g_0^2\bigg(\overline\psi \gamma^\mu\,\gamma^5\,T^b\psi\bigg)^2
%\bigg(\overline\psi \gamma_\mu\,\gamma^5\,T^c\psi\bigg)
%\nonumber\\
\eea
with twice the squared screening mass observed in (\ref{9X}).
The color anomaly fixes the gluon operator contribution in (\ref{3X}).
The same result readily follows in the
center gauge, and therefore holds for all gauges by gauge invariance.
Most of the identities discussed in this section simplify for Abelian QED or the Schwinger model, thanks again to
the U(1) axial anomaly. For completeness, they are quoted in Appendix~\ref{QED}.
As the first application of (\ref{10}), let us derive the vacuum gluon condensate

\bea
\label{10X}
\left<\bigg(\partial^2+\frac {m_0^2}{N_c}\bigg) (E^aE^a)\right>=\frac{m_0^2}{N_c}\left<E^aE^a\right>=
2g_0^2\left<\bigg(\overline\psi \gamma^\mu\,\gamma^5\,T^b\psi\bigg)^2\right>
\eea
The colored 4-Fermi current-current contribution can be simplified  into colorless interactions by  Fierzing.
% by using the color identity
%\be
%\label{11}
%(T^c)^{ij}(T^c)^{kl}=2\bigg(\delta^{il}\delta^{jk}-\frac 1{N_c}\delta^{ij}\delta^{kl}\bigg)
%\ee
%and Fierzing, so that
%\be
%\label{11X}
%\left<E^aE^a\right>=2\pi\bigg<\bigg(\overline\psi \gamma^\mu\,\gamma^5\,T^c\psi\bigg)^2
%=2\bigg(\big(\overline\psi \psi\big)^2+\big(\overline\psi i\gamma^5\psi\big)^2-\frac 2{N_c}\big(\overline\psi \gamma^\mu\psi\big)^2\bigg)\bigg>
%\ee
%with manifest chiral symmetry as it should.
Assuming vacuum factorization which is justified in large $N_c$ by the master
field~\cite{Witten:1979pi}, we can  simplify the gluon condensate to

\be
\label{EEVAC}
\left<E^aE^a\right>=2\pi\bigg<\bigg(\overline\psi \gamma^\mu\,\gamma^5\,T^c\psi\bigg)^2\bigg>
=\pi\frac{N_c^2-1}{N_c^2}\left<\overline\psi\psi\right>^2\rightarrow \pi\left<\overline\psi\psi\right>^2
\ee
with the rightmost result in agreement with the large $N_c$ result  briefly quoted  in~\cite{Zhitnitsky:1985um}. The fact that the quark and gluon
condensates are tied in two dimensions, is reminiscent of the way   both condensates are tied  to the instanton packing
fraction in four dimensions~\cite{Schafer:1996wv,Nowak:1996aj}.

We now  use (\ref{10}) to characterize a dual  mass sum rule for the bound mesons. The gluon mass operator  in
Eq.~(\ref{Trace}) can be unwound in terms of  4-Fermi colored interactions much like in the vacuum in (\ref{EEVAC}), with the result

\bea
\label{MASSX}
M_n^2=
%&&-\pi\left<P|(\overline\psi \gamma^\lambda\,\gamma^5\,T^a\psi)^2|P\right>+\frac 12\left<P\left|m\overline\psi\psi\right|P\right>\nonumber\\
%=&&
2\pi\left<P,n\left| (\overline\psi \psi\big)^2+\big(\overline\psi i\gamma^5\psi\big)^2-\frac 2{N_c}\big(\overline\psi \gamma^\mu\psi\big)^2\right|P,n\right>
+\frac 12\left<P,n\left|m\overline\psi\psi\right|P,n\right>
\eea
\end{widetext}
after Fierz rearrangement. In two-dimensions, the gluon field is longitudinal and sourced
by the quarks  in the bound state. The fermionic mass sum rule (\ref{MASSX}) is dual to the gluonic mass sum rule (\ref{SRIMF}),
and it is somehow illusive to try to distinguish one
from the other. This is reminiscent of the Coulomb energy between a pair of classical charges which can be either assessed
using (screened) Coulomb law between the charges, or the energy stored in the (screened) Coulomb field  surrounding the charges. In most of
the above discussions, this should be kept in mind.

We  note that (\ref{MASSX}) is how the meson masses are generated in models that capture chiral symmetry breaking {\it without} confinement,
such as the instanton model for the QCD vacuum~\cite{Schafer:1996wv,Nowak:1996aj},
and bound state models~\cite{Nowak:1996aj,Bashir:2012fs,Eichmann:2016yit}
 in four dimensions.  For the former, the strong instanton and anti-instanton fields are
able to trap quarks into left- and right-handed zero mode states, thanks to the axial U(1) anomaly,  and generate quasi-local 4-Fermi determinantal interactions. The latters
are at the origin of finite constituent masses and meson masses much like (\ref{MASSX}). It is remarkable, that in two dimensions and large $N_c$, the subleading
quantum  screening of the  electric field  as enforced by  the  colored axial-vector  anomaly (\ref{7}),
is also at the origin of  the leading chirally symmetric and local 4-Fermi interactions,  with a dual mass relation for the bound mesons in any frame. In  a way,
the spontaneous breaking of chiral symmetry appears to be the chief non-perturbative mechanism for mass generation for the light hadrons, in a confining theory such as QCD where the string breaks through quantum screening by light quarks.

\section{Conclusions}\label{CONCLUSIONS}

We have explored the role of the chiral and gluon matrix elements in
the composition of the mass of the meson bound states in two dimensional QCD in the $^\prime$t Hooft limit.
Both the chiral and gluon matrix elements in  a bound meson are probed by the chiral-odd quark parton distribution,
and the gluon parton distribution respectively.  We have explicitly derived these distributions in the $^\prime$t Hooft limit. Their zeroth moments are indeed related to the quark and gluon condensate in the meson states. For the chiral odd quark
distribution, we have shown that the lowest moment arises not from a delta-function  accumulation of wee partons at the edge or $x\sim 0$, but rather from
their rapid and specific depletion at the edge. For the gluon distribution, the situation is more subtle. Color neutrality forces the gluon distribution $xF_n(x)$
to be finite as $x\rightarrow 0$ without the occurence of a singular contribution, yet the mass sum rule in the chiral limit is fulfilled.
Our use of the PV prescription does not upset  the light front sum rules in leading order in $1/N_c$.

The form factor of the energy-momentum tensor splits into two invariant form factors, one for the
graviton or tensor coupling, and the other for the dilaton or scalar  coupling. However,
in two dimensions they are related. The tracefull
part is  similar to that in four dimensional QCD,  although for different reasons since
two-dimensional QCD is non-conformal. The tracefull part of the energy
momentum tensor fixes the bound meson masses, and dominates the  gravitational
form factor. Finally but remarkably, a dual mass sum rule is shown to emerge solely
in terms of chirally invariant 4-Fermi interactions.

The present results may be specific to two dimensions, since QCD  is non-conformal.  Also, in the light cone gauge,
the structure of the light cone zero modes arising from a perturbative treatment, is regulated by
quantum mechanics in the absence of the transverse degrees of freedom. In four dimensions these zero modes are regulated by field theory,
and may be at the origin of the singular contributions in some higher
twist contributions to the parton distribution functions, as we discussed. More work in this direction is needed.

\section{Acknowledgements}

This work was supported by the U.S. Department of Energy under Contract No. DE-SC0020682 for XJ and DE-FG-88ER40388  for IZ.
\vskip 1cm

%\newpage

\appendix

\section{Massive Abelian QED}\label{QED}

The analysis in section~\ref{ANOMALYX} carries to two dimensional massive QED with more simplifications.
The U(1) axial anomaly Abelianizes

\be
\label{7Q}
\partial_\mu\bigg(\overline\psi \gamma^\mu\,\gamma^5\psi\bigg)=\frac {e}{2\pi}\epsilon_{\mu\nu}F^{\mu\nu}= \frac{e}{\pi}E
\ee
which is again given by the one-loop diagram.  Following the same reasoning as given earlier,  we have for Maxwell equation

\be
\label{5Q}
\partial_\mu F^{\mu\nu}=-e\overline\psi \gamma^\nu\psi
\ee
and using the duality of the currents

\be
\label{9Q}
\partial_\mu E=-e\overline\psi \gamma_\mu\,\gamma^5\psi
\ee
This shows that the pseudo-potential $E$ describes a massive free boson

\be
\label{9XQ}
\bigg(\partial^2+\frac{e^2}{\pi}\bigg)E=0
\ee
with $E^2$ obeying the screened source equation

\bea
\label{10Q}
\bigg(\partial^2+\frac {2e^2}{\pi}\bigg) E^2
=
%=2(D^2E^a)E^a+2(D_\mu E^a)D^\mu E^a)\nonumber\\
2e^2\bigg(\overline\psi \gamma^\mu\,\gamma^5\psi\bigg)^2=-2e^2\bigg(\overline\psi \gamma^\mu\psi\bigg)^2\nonumber\\
%\bigg(\overline\psi \gamma_\mu\,\gamma^5\,T^c\psi\bigg)
%\nonumber\\
\eea
with again twice the squared screening mass.
In the massive case,
the squared meson mass obeys the analogue of the sum rule (\ref{MASS}), namely

\begin{widetext}

\be
\label{MASS2}
M^2=\frac 12\left<P\left| E^2\right|P\right>+\frac 12 \left<P\left|m\overline\psi\psi\right|P\right>=
\frac{\pi}2 \left<P\left| \left(\overline\psi\gamma^\mu\gamma^5\psi\right)^2\right|P\right>+\frac 12 \left<P\left|m\overline\psi\psi\right|P\right>
\ee
In two dimensions the currents bosonize

\be
\label{BOSO}
\overline \psi i\gamma^\mu\gamma^5\psi=\frac 1{\sqrt\pi}\partial^\mu\phi\qquad\qquad
\overline \psi\gamma^\mu\psi=-\frac 1{\sqrt\pi}\epsilon^{\mu\nu}\partial_\nu\phi
\ee
In the massless case $\phi$  is a free massive boson with  squared mass $M^2=e^2/\pi$, which is seen to be satisfied by (\ref{MASS2})

\be
M^2\rightarrow \frac 12\left<P\left| E^2\right|P\right>=-\frac{\pi}{2(\sqrt\pi)^2}\left<P\left|(\partial_\mu\phi)^2\right|P\right>
=M^2\bigg[\frac 12\left<P\left|\phi^2\right|P\right>=1\bigg]
\ee
The last bracket as a scalar is frame invariant. We evaluated it in the meson rest frame,  using   the equal-time free field decomposition

\be
\phi(x)=\int\frac{dk}{2\pi}\frac 1{\sqrt{E_k}}\left(e^{-ik\cdot x}\alpha(k)+e^{ik\cdot x}\alpha^\dagger(k)\right)
\ee
and the canonical rules $\left[\alpha(k), \alpha^\dagger(p)\right]=2\pi\delta(k-p)$. The meson state with scattering normalization is defined as
$\left|P\right>=\sqrt{P^0}(\alpha^\dagger(P)\left|0\right>)$ with $P^0=E_P=M$ at rest.
\end{widetext}

%Therefore in the following discussions, we will always keep this point in mind.

\section{Behavior of Light-Front Wave Function Near the Edges in the Chiral Limit}\label{LIGHT}

The wavefunctions solution to the $^\prime$t Hooft equation Eq.~(\ref{THOOFT}), describe
the quark parton distribution amplitudes on the light cone

\begin{align}\label{PN}
\varphi_n(x)=\frac {1}{f_0}\int \frac{dz}{2\pi}e^{ixP^+z}\left<0\left|\overline\psi_+(0)[0,z]\psi_+(z)\right|P,n\right>
\end{align}
in the  gauge $A_-=0$, with $[0,z]=1$.  They satisfy
conjugate symmetry  $\varphi_n(1-x)=(-1)^n\varphi_n(x)$ along with a number of integral identities~\cite{Callan:1975ps}.
They are orthonormal

\be
\int_0^1dx \varphi^*_n(x)\varphi_n(x)=\delta_{nm}
\ee
and vanish at the edges

\be
\label{EDGE}
\varphi_n(x\sim 0,1)\sim C_n(x\bar x)^\beta
%\qquad\qquad x\sim 0,1
\ee
with $\beta$ solution to

\be
\pi\beta{\rm cot}(\pi\beta)=\frac{\beta^2}{m_0^2}-\frac{m^2}{m_0^2}
\ee
which for small masses give $\beta\sim \sqrt{3}m/{\pi m_0}$. Away from the edges, the solutions
are not known analytically except for the almost Goldstone mode with $M_0\sim {\cal{O}}(m)$, $C_0=1$
and $\varphi_0(x)=\theta(x\bar x)$ (except at the edges)~\cite{Zhitnitsky:1985um}. In general, the exact solutions to (\ref{THOOFT}) are only
known numerically. For large $n$, the solutions follow analytically from semi-classics
with $\varphi_n(x)\sim \sqrt{2}{\rm sin}((n+1)\pi x+\delta_n(x))$ (except at the edges) and $M_n^2\approx n\pi^2 m_0^2$~\cite{tHooft:1974pnl,Callan:1975ps}.
The  pseudoscalar transition amplitude
%\begin{widetext}
\be
\label{DECAY}
\left<0\left|\overline\psi i\gamma^5\psi\right|P,n\right>
%=\sqrt{\frac{N_c}{\pi}}\pi m_0f_n
=\sqrt{2}f_0\bigg[\frac m{2}\int_0^1dx \frac{\varphi_n(x)}{x\bar x}\bigg]\nonumber\\
\ee
%\end{widetext}
fixes  the decay constant $f_0=({N_c/2\pi})^{\frac 12}$.

\section{Virial theorem}\label{virial}

In this Appendix we provide a simple derivation of the Virial theorem ${\cal K}={\cal G}$ established in Eq.~(\ref{eq:K=G}). 
 For that, consider the rescaling of the only dimensionful parameters
 $g_0^2=M^2\lambda_1$ and $m=M\lambda_2$ with $M$ being a common mass parameter and $\lambda_1$, $\lambda_2$ being massless.  All  meson 
 masses are proportional to $M$ by dimensions, i.e. $M_n=Mf(\lambda_1,\lambda_2)$. Using the Feynman-Hellman theorem

\begin{align}
M_n=\langle M\frac{\partial H}{\partial M}\rangle \ ,
\end{align}
we have

\begin{align}
M_n={\cal M}+2{\cal G} \ ,
\end{align}
since the gluon energy in the Hamiltonian (\ref{eq:H}) is propagational to $g_0^2\propto M^2$ and the mass term is proportional to $m \propto M$. By comparing with the mass decomposition,
\begin{align}
M_n={\cal M}+{\cal G}+{\cal K} \ ,
\end{align}
we found
\begin{align}
{\cal G}={\cal K},
\end{align}
which is the desired virial theorem.

In the large-$N_c$ limit, the Hamiltonian in Eq.~(\ref{eq:H}) can be expressed through bosonization in terms of meson operators~\cite{Jia:2018qee}.  The Bars-Green equation arises as one tries to diagonalise the Hamiltonian. By applying the Virial theorem to the bosonized form of the Hamiltonian and then expressing in terms of the components, one obtains the desired virial theorem expressed in terms of the Bars-Green wave functions.

\bibliographystyle{apsrev4-1}
\bibliography{bibliography}

\end{document}